\documentclass{emulateapj}

\usepackage{subfigure}
 \usepackage{graphicx}
\usepackage{threeparttable}
\usepackage{rotating}
\usepackage{lipsum}
\usepackage{arydshln}
\usepackage{amsmath}
\usepackage{amssymb}
\usepackage[ ampersand]{easylist}
\usepackage{verbatim}
\usepackage{color, soul}
\usepackage{calc}

\usepackage{array}
\usepackage{booktabs}
\newcommand{\lya}{Lyman-$\alpha$}
\newcommand{\wlya}{\hbox{$W_{\mathrm{Ly}\alpha}$}}

\newsavebox\CBox
\newcommand\hcancel[2][0.5pt]{%
  \ifmmode\sbox\CBox{$#2$}\else\sbox\CBox{#2}\fi%
  \makebox[0pt][l]{\usebox\CBox}%
  \rule[0.5\ht\CBox-#1/2]{\wd\CBox}{#1}}

  \newcommand{\ergscm}{$\rm erg \; s^{-1}\; cm^{-2}$}

   \newcommand{\muv}{$M_{UV}$}

\newcommand{\ep}{$\epsilon_{ne}$} 
\newcommand{\es}{$\epsilon_{de}$} 

\def\ppd{\ifmmode {\Pp_d}\else
               ${\Pp_d}$\fi}

\newcommand{\fhi}{$f_{\rm HI}$ }

\newcommand{\ed}[1]{\textcolor{black}{{#1}}}
\newcommand{\cjp}[1]{\textcolor{black}{{#1}}}

\slugcomment{}

\shorttitle{Rapid Decline of  \lya\ Emission Toward The Reionization Era
}
\shortauthors{Tilvi et al.}

\begin{document}

\title{Rapid Decline of  LYMAN-$\alpha$ Emission Toward the Reionization Era
}


\author{V.  Tilvi  \altaffilmark{1},
C. Papovich\altaffilmark{1},
S. L. Finkelstein\altaffilmark{2},
J. Long\altaffilmark{3},
M. Song\altaffilmark{2},
M. Dickinson \altaffilmark{4},
H. C. Ferguson\altaffilmark{5}, 
A. M. Koekemoer\altaffilmark{5},
M. Giavalisco\altaffilmark{6}, and 
B. Mobasher\altaffilmark{7}
 }

\altaffiltext{1}{George P. and Cynthia Woods Mitchell Institute for Fundamental Physics 
and Astronomy, and Department of Physics and Astronomy, Texas A\&M University, 
College Station, TX. 77845}
\altaffiltext{2}{Department of Astronomy, The University of Texas at Austin, Austin, 
TX. 78712}  
\altaffiltext{3}{Department of Statistics, Texas A\&M, College Station, TX. 77845}  
\altaffiltext{4}{National Optical Astronomy Observatory, Tuscon, AZ. 85719}
\altaffiltext{5}{Space Telescope Science Institute, Baltimore, MD. 21218}  
\altaffiltext{6}{Department of Astronomy, University of  Massachusetts, Amherst, 
MA. 01003}  
\altaffiltext{7}{Department of Physics and Astronomy, University of California, Riverside, 
CA. 92521 }
\altaffiltext{*}{The data presented herein were obtained at the W.M. Keck Observatory, 
which is operated as a scientific partnership among the California Institute of Technology, 
the University of California and the National Aeronautics and Space Administration. The 
Observatory was made possible by the generous financial support of the W.M. Keck 
Foundation.}

\begin{abstract} 
The observed deficit  of  strongly \lya\ emitting galaxies   at
$z>6.5$ is attributed to either   increasing neutral hydrogen  in the
intergalactic medium (IGM) and/or  to the evolving galaxy
properties.  To investigate this,  we have performed  very
deep near-IR spectroscopy of $z\gtrsim7$ galaxies using MOSFIRE on
the Keck-I Telescope.  We measure the \lya\ fraction at
$z\sim8$  using  two methods.  
First, we derived
$N_{\mathrm{Ly}\alpha}/N_\mathrm{tot}$ directly using extensive
simulations  to correct for incompleteness.  Second, we  used a
Bayesian formalism  (introduced by Treu et
al. 2012) that compares the $z>7$ galaxy spectra to models of
the \lya\ equivalent width (\wlya) distribution at  $z\sim6$.
We explored two simple evolutionary scenarios: 
 pure \textit{number evolution}
 where \lya\ is blocked in some fraction of galaxies (perhaps
due to the IGM being  opaque along only some fraction of  sightlines) and 
  uniform \textit{dimming evolution} 
 where   \lya\ is attenuated
in all galaxies by a constant factor (perhaps owing to processes from
galaxy evolution or a slowly increasing IGM opacity).  
  The
Bayesian formalism places stronger constraints compared with the
direct method.  Combining our data with that in the literature we find
that at $z\sim8$ the  \lya\ fraction has dropped by a
factor $>$3 (84\% confidence interval) using both  the 
dimming 
and
number evolution scenarios, compared to the $z\sim6$ values.  
Furthermore, we find a tentative ``positive'' Bayesian evidence 
 favoring the number evolution scenario
  over  dimming evolution, extending trends
  observed at $z\lesssim7$ to higher redshift.
Comparison of our results  with theoretical models  imply the IGM volume 
averaged neutral hydrogen fraction  
 $\gtrsim0.3$
 suggesting that we are likely witnessing  reionization in progress at  $z\sim8$. 
  \end{abstract}

\section{Introduction} 
With  growing  number   of  spectroscopically
confirmed  galaxies  at  $z>6.5$, it is evident  that there is  a
dearth of galaxies with high rest-frame \lya\ equivalent widths
(\wlya).  We illustrate this problem in  Figure~1, showing the
observed   $W_{\mathrm{Ly}\alpha}$ for  galaxies with high spectroscopic
confidence
at $z > 6.5$
\citep{iye06, ouc10, rho12, fin13, pen14}.
The lack of high-$W_{\mathrm{Ly}\alpha}$ galaxies  is unlikely  due
to selection bias as these galaxies  span a wide range of continuum
magnitudes (lower  panels in Figure~1) i.e.,   we are not just limited to 
some brighter UV continuum galaxies causing the observed  decline
in  \wlya.
  A similar trend is also
observed in the \lya\ fraction of continuum-selected Lyman-break
galaxies:  while the fraction of \lya\  galaxies increases from $z=3$
to 6 \citep{sta11}, there is a marked decline at $z>7$  \citep{fon10, rob10,
van11,  ono12, sch12, car12,  tre13, pen14, fai14, sch14}.

 \begin{figure*}[t]
\epsscale{1}
\plotone{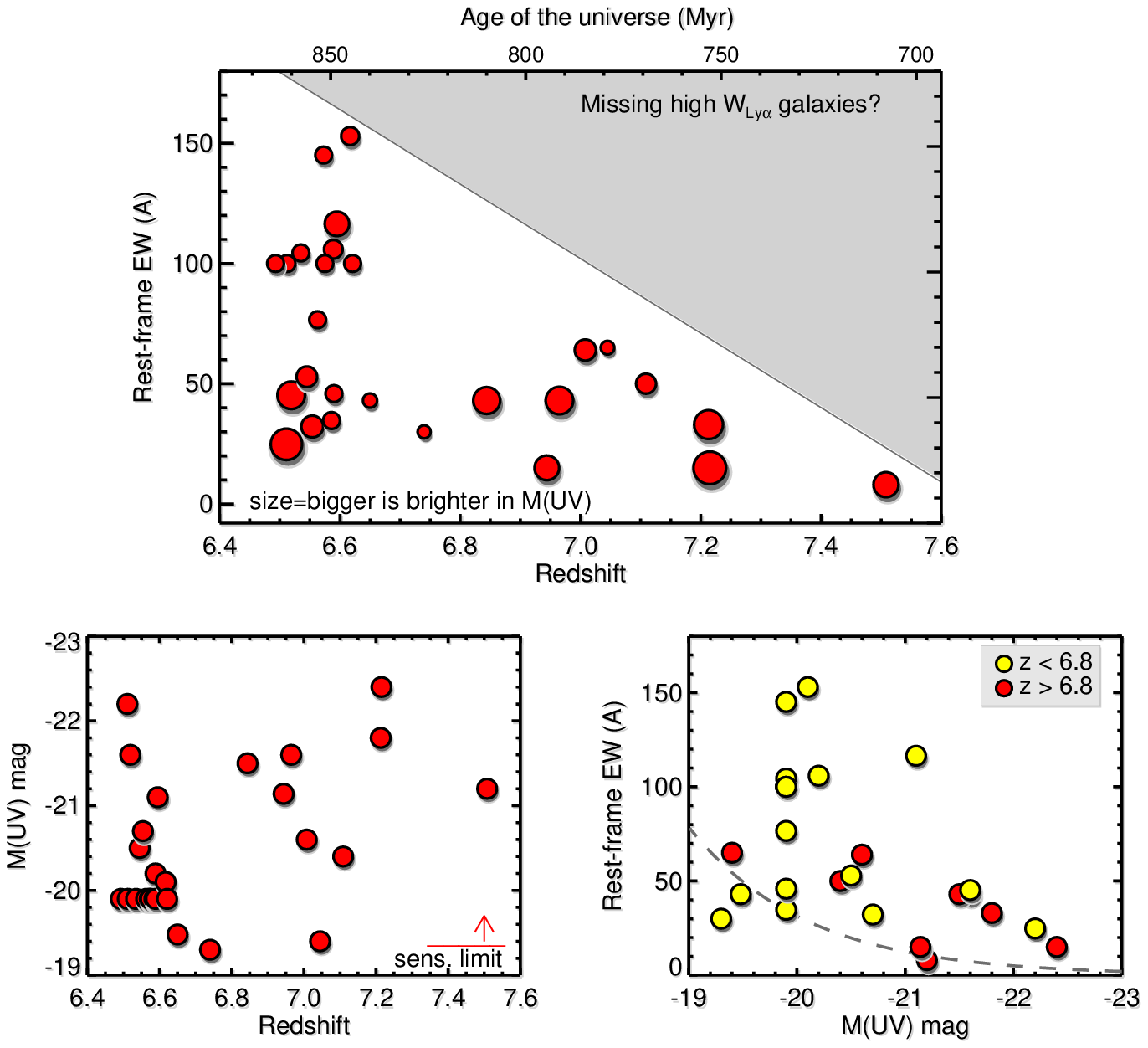}
\caption{
\textbf{Missing high \lya\ equivalent width galaxies} $-$Top  panel:
Redshift evolution of rest-frame \lya\ equivalent width (\wlya) for 
galaxies with high spectroscopic confidence 
\citep{iye06, ouc10, sch12, van11, ono12, rho12, fin13, pen14}.
There is a missing population of high \wlya\ galaxies at $z\gtrsim 7$.
This deficit is not due to  selection bias or observing galaxies with only 
brighter continuum 
magnitudes (lower left panel) as these galaxies span a wide range of 
\muv\ magnitudes. 
The clustering of galaxies at \muv=-20 is due to the survey limit.
Bottom-right panel shows that the galaxies at $z<6.8$ as well 
as  $z>6.8$ span a
similar range of  \muv\ magnitudes. The dashed line shows a typical 
$z\gtrsim7$ spectroscopic survey  limit, assuming a \lya\ line flux limit
$5\times 10^{-18} \rm erg \;s^{-1} cm^{-2}$; Treu et al (2013), Finkelstein 
et al (2013)   : current spectroscopic surveys at $z>7$ are sensitive to 
galaxies with \wlya\ greater than this threshold.
\vspace{0.3cm}
}
\end{figure*}

Clearly something is changing in the \lya\ emitting population at
$z\gtrsim7$. 
As \lya\ emission is sensitive to  neutral  hydrogen fraction in the 
intergalactic medium \citep[IGM:][]{mcq07},
it is tempting to associate the decline in the \lya\
fraction with an increasing neutral hydrogen  fraction in the
IGM, as inferred from QSO sightlines  at these
redshifts \citep{fan06} because there is no indication that the
galaxy properties contributing to   \wlya\ 
 are evolving rapidly.   For example,
at $3<z<6$ \citep{ouc08,sta11,mal12, zhe14}, where the IGM
 is mostly ionized \citep{fan02}, there is no observed evolution
  in the \wlya\ distribution and this 
offers insight into the
galaxies' physical processes.
Also, there is no evolution in the  number density of \lya\ emitting
galaxies  in this redshift range \citep{kas06,iye06,daw07, ouc08}.
Clearly, if  it were known that the (intrinsic)
$W_{\mathrm{Ly}\alpha}$ distribution at $3<z<6$ continues to higher
redshift, then the observed decline in $W_{\mathrm{Ly}\alpha}$ at
$z>7$ must stem from an increasing  neutral hydrogen fraction in the
IGM. 
\cjp{For example, \citet{kon14} recently reported a marked decline
in the number density of \lya-emitting population at $z\sim 7.3$ from
narrow-band imaging, consistent with a declining $W_\mathrm{Ly\alpha}$.
}

The declining \wlya\ distribution could suggest an increasing neutral hydrogen 
fraction with redshift.  This is consistent with recent theoretical studies 
\citep[e.g.,][]{for12}
that the observed decline in the
\lya\ fraction at $z\sim7$ requires  about  $\sim10-20\%$ neutral
hydrogen, when combined with field-to-field variance \citep{tay13}, a
possibly evolving escape fraction of ionizing photons \citep{dij14},
and/or incidence of \lya\ absorption systems \citep{bol13}.  
\cjp{If reionization is in fact extremely rapid, with the neutral fraction
evolving from $>$10\% to $< 0.01$\%  in the $\lesssim$200 Myr from
$z\sim7$ and  $z\sim6$, then one would indeed expect a strong
evolution in the number of \lya\ galaxies formed in this short
redshift range \citep{jen13}.}

\begin{figure*}[t!]
\epsscale{1.1}
\plotone{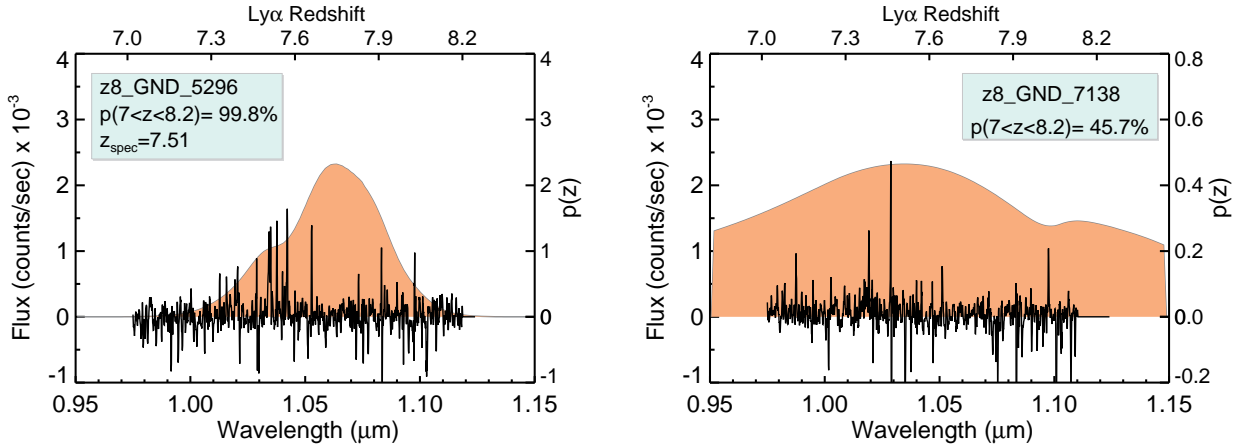}
\caption{
MOSFIRE Y-band spectra  of two galaxies, representative of our sample.
The shaded  region shows the photometric probability density $p(z)$ 
for each galaxy.  The left panel shows the object
that was spectroscopically confirmed with \lya\ line detection at
$z=7.51$ \citep{fin13},
 with a narrow $p(z)$.  The
right panel shows an object with no line detection, and a broad
$p(z)$.
While deriving the \lya\ fraction, we use all available information in the 
1D spectra, as well as correcting our analysis for $p(z)$ outside of our 
observed spectroscopic wavelength range.
\vspace{0.3cm}
}
\end{figure*}

On the other hand, the evolution in \lya\ emission may signify
evolution in the physical properties of galaxies at $z > 6$.  
 And
indeed, some recent works at $z>7$ show some evolution in galaxy's
physical properties.   Galaxies at $z>7$ have bluer colors
\citep[][but see also McLure 2011, Dunlop et al 2012]{bou10b, lab10,
wil11, fin12, til13}, likely due to lower dust extinction,
\citep[e.g.,][]{fin12}, with lower stellar mass \citep{fin10, sch10},
and smaller sizes \citep{malh12, ono13}.  However, these results, if
anything, should support the idea that the decline  in \wlya\ and
\lya\ fraction is caused by the increasing neutral hydrogen
fraction; lesser dust and smaller masses would make it easier for
\lya\ photons to escape unless it is a result of gas accretion
\citep{ker09} with high covering fraction \citep[but see also][]{jon13}.  
 \cjp{Indeed, this may be the case as
empirical arguments suggest that   the  gas accretion rate
exceeds the star-formation rate at $z \gg 4$ \citep{pap11}.}
Therefore, there are plausible reasons to suspect that any evolution in 
the UV continuum properties of LBGs at $\gtrsim7$ might also 
contribute to the evolution in \wlya\citep[e.g.,][]{fin12,lor13, bou14}.

In this  paper, 
we measure the redshift evolution of \lya\ emission at $z > 7$ and  we
study the nature of the evolution of \wlya\ using simple empirical models.
In addition to using our deep spectroscopic observations \citep{fin13},
to increase the sample size in order to mitigate the effects of cosmic 
(field-to-field) variance
\citep{til09}
and increase the 
significance from independent datasets,
we also combine our
data with observations from the literature \citep{tre13}.
We measure the evolution of the \lya\  
fraction (the fraction of galaxies with $W_{\mathrm{Ly}\alpha}$ 
above a certain limit)
using both a direct measurement of 
$N_\mathrm{Ly\alpha}$/$N_\mathrm{tot}$ (\S 3.1), and testing 
the $z>7$ 
$W_\mathrm{Ly\alpha}$ distribution against that at $z\sim6$  using
a Bayesian formalism  
against an empirical model
\citep[\S~3.2;][]{tre12, tre13}.
In \S 4 and \S 5 we discuss our results and present a summary of our 
findings, respectively.  Where applicable, we assume cosmological 
parameters $\Omega_M = 0.27$, $\Omega_\Lambda = 0.73$, 
and $H_0 = 71$ km s$^{-1}$ Mpc$^{-1}$.Ó

\begin{figure*}
\epsscale{1.18}
\plotone{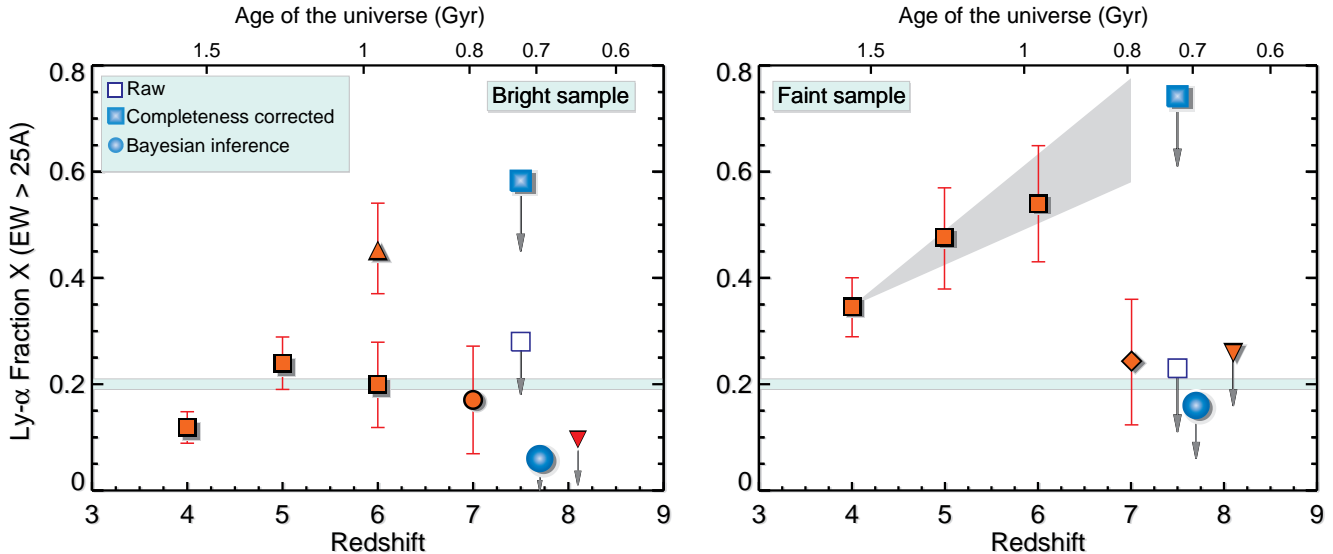}
\caption{
Redshift evolution of the \lya\ fraction with  $\wlya>25$~\AA.
The left and right panels show results for brighter ($M_\mathrm{UV}<-20.25$)
and  fainter sample ($M_\mathrm{UV}>-20.25$), respectively.
The blue open square indicates  the uncorrected limit (for our data only; 84\% confidence) 
on the
fraction from
the direct method (\S~3.1) while the filled square shows the limit 
including completeness corrections and accounting for  the
photometric redshift distributions.   The filled blue circle shows the
\lya\ fraction  derived using the Bayesian
formalism (\S~3.2).  
Red filled squares, upward-pointing triangle,  and downward-pointing 
triangles are
data from \citep{sta11, cur12, ono12}, respectively. The filled diamond 
is the combined
data taken from \citet{ono12} which is composed of data from 
\citep{fon10, pen11, sch12}.
There is a  difference in the \lya\ fraction estimated from the direct method and Bayesian method
which stems from using our data only and using the combined data (with Treu et al 2013), respectively.
But the unique advantage of using the Bayesian inference is that the results at $z\sim8$ 
are \textit{relative} to the $z\sim6$ distribution.
  If the drop in the \lya\
fraction is due to increasing neutral hydrogen fraction in the IGM,
this occurs over a short, $<$300 Myr, period, and we are likely
witnessing  reionization in progress at $z>7$. 
\vspace{0.3cm}
}
\end{figure*}

\section{Candidate Selection and Spectroscopic Observations} 
 
To select $z\gtrsim7$ candidate galaxies  for spectroscopic follow-up we 
used extremely deep
 WFC3/F160W- selected and PSF-matched photometric catalog \citep{fin13} 
 created using
 imaging from 
 the Cosmic Assembly Near-infrared Deep Extragalactic Legacy Survey 
 \citep[CANDELS: ][]{gro11, koe11}.
All candidates were selected using 
photometric redshifts,  derived using the photometric redshift code 
EAZY \citep{bra08}
which  uses redshifted spectral energy distribution templates and
 compare them with the 
observed multi-band photometry for a given galaxy. In addition to 
deriving the best-fit 
photometric redshift, it provides the photometric redshift probability 
density $p(z)$ for each
object. In the following analyses we make use of full photometric 
redshift distribution
instead of just the best-fit  redshifts.
Each object in our sample was required to have  
$>70\%$ of the integral of $p(z)$ in the primary peak and $ >25\%$  
 at $z=7.5-8.5$ for the $z\sim8$ sample and at $z=6.5-7.5$ for 
 the $z\sim7$ sample, 
and be detected in both F125W and F160W bands at S/N$>3.5$.
For further details about the candidate selection and data reduction and 
calibration, we 
direct the reader to Finkelstein et al. (2013).

We  targeted  
nine $z\sim8$ and 34 $z\sim7$  candidate  galaxies  using Multi-object 
Spectrometer For Infra-Red Exploration
\citep[MOSFIRE:][]{mcl12}, a
near-IR multi-object spectrograph   on the Keck Telescope,   from Apr
17-18, 2013.  
The $z\sim8$ selection of objects to be put on MOSFIRE masks 
 were prioritized  based on their magnitude and the amount of their $p(z)$ 
contained within the redshift range $7<z< 8.2$, covered by the  
MOSFIRE Y-band.

In this study, we focus  on nine $z\sim8$ galaxies that have 
spectroscopic observations.
 The spectroscopic observing conditions were excellent with median 
 seeing FWHM$\simeq 0\farcs7$.
We used the MOSFIRE data reduction pipeline to reduce the raw data 
and extract 2D, wavelength
calibrated spectra. 
To flux calibrate the 1D spectra, we used the standard star 
measurements
taken during the same observing nights. We  found that the flux errors 
in the reduced 1D 
spectra are overestimated. In order to correct these errors, we used the 
ratio of standard deviation
of flux and median value of flux errors for a given spectrum, and scaled the
 flux errors such
that this ratio is close to unity (as was done in \citealt{fin13}). These 
corrected flux errors are consistent with the MOSFIRE 
exposure time calculator. 
The typical exposure time per galaxy is about 5--6 hours which allowed us 
to reach 
deep line flux sensitivity of $\simeq$$2\times 10^{-18}$~\ergscm\ (5$\sigma$: 
although this limit varies
with wavelength and the presence of sky emission features).

Our MOSFIRE observations used the $Y$-band filter covering 0.97-1.12 
\micron m, sensitive to 
\lya\ emission from galaxies at $7 < z < 8.2$.  Figure~2 shows spectra 
of two galaxies 
typical of our sample, along with their photometric redshift probability 
densities, $p(z)$.  
The galaxy  in the left panel of Figure~2 has a narrow $p(z)$ with 
$\mathcal{P}_z \equiv \int_{z=7.0}^{z=8.2}\ p(z)\ dz=99.8\%$, 
while the galaxy in the right panel has a broad $p(z)$ with 
$\mathcal{P}_z=45.7$\%.

\section{Redshift Evolution of  Lyman-$\alpha$ Fraction} 

\subsection{Direct Measure} 

The most straightforward measure of the \lya\ fraction  is the ratio
of number of  galaxies with \lya\ emission to the total number of
galaxies observed.  Figure~3 shows the redshift evolution  of \lya\
fraction with   $\wlya>25$~\AA.   \cjp{Formally, none of the objects in
our sample are detected with \wlya\ greater than this value and so 
our measurements
of the fraction are upper limits.}  The
open blue squares show the raw \lya\ fraction obtained from our deep
MOSFIRE observations, 
\cjp{$X=N_{Ly\alpha}/N_{tot}<0.28$ 
(84\% confidence derived from Poisson statistics for 0/4 objects  
with EW$>$25\AA)
for the
brighter ($M_\mathrm{UV}<-20.25$~mag)  sample  and $X<$0.23 
(84\% confidence for 0/5 objects) for 
the fainter  sample ($M_\mathrm{UV}>-20.25$~mag).}

The raw fractions above must be corrected for the wavelength dependent 
sensitivity of the observed spectra.
To estimate this correction,  we performed extensive simulations, similar 
to those described in
\citet{sta11}.
First, we inserted artificial  sources at random positions along the
dispersion direction in the reduced  2D spectra.
We then recover the
sources using the same automated method as for the real data: 
extracting one-dimensional spectra, fitting a Gaussian profile to
significant lines and then compute the S/N from the fit.   Emission
features with S/N $>5$ are considered ``recovered''.
The completeness $C_i^\prime(m,W,z)$ for an object is then 
\begin{equation}
C_i^\prime(m_i,W_i, z)=\rm N_{rec}/N_{ins},
\end{equation}
where $ \rm N_{ins}$ and $\rm N_{rec}$ are the number of simulated
(inserted) and recovered artificial sources in the 2D
spectra at a given apparent magnitude, $m_i$,  with \lya\ equivalent 
width
$W_i$, and redshift $z$. 

We must modify the completeness function to account for $p(z)$, 
because there is 
a probability of  Lyman-$\alpha$ falling  outside the wavelength range
 covered 
by the $Y$-band filter for some objects, and thus

\begin{equation}
C_i(m_i,W_i)=\int\limits_{z=7}^{z=8.2}C_i^\prime(m_i,W_i, z)\; p(z)\; dz 
%
%
\end{equation}
The effective completeness for all objects  with  equivalent
 width W 
and   absolute magnitude  M (M=$m -\mu$, where $\mu$ is
 the distance
modulus and m is the observed F125W magnitude)  is then
\begin{equation}
 C_{\rm eff}(M,W)=\displaystyle \sum\limits_{i=1}^N  C_i(m_i,W_i)/N, 
\end{equation}
where $N$ is  the total number of galaxies in the given (brighter or
 fainter) sample.
The
\lya\ fraction corrected for incompleteness   is then 
$X_{\rm corr}=X/C_{\rm eff}$.

In Figure~3 filled squares  show the completeness corrected \lya\ 
fraction.   
The completeness corrected fraction  is \cjp{$X_{\rm corr}$ 
$<$ 0.58}
for the brighter subsample, and \cjp{$<$}0.74 for the fainter sample. 
  For our
sample, this is primarily a 
consequence of the fact that the $p(z)$ is not  entirely covered 
by the
 MOSFIRE $Y$-band  observations.
 On average
$\langle \mathcal{P}_z \rangle= 0.45$,
 and this has the effect of
nearly doubling the \lya\ fraction.    For this reason, in the following 
sections, we explore
an alternative method to measure $X$.

\subsection{Bayesian Inference}

We use a Bayesian  formalism developed by \citet{tre12} to 
measure the
evolution of the \lya\ equivalent width distribution from $z\sim6$ 
to $z>7$.
This method constrains the \lya\ equivalent width
distribution based on all available information;  detections,
non-detections, wavelength-dependent line-flux sensitivity, and
incomplete wavelength coverage (similar to the Direct method with 
completeness simulations, \S 3.1).   An  advantage of this 
formalism  is that the
results obtained from different data sets or instruments can be
combined together easily by simply multiplying the posterior
probabilities, and that the probability at $z>7$ is  \textit{relative} to 
the $z\sim6$ 
distribution; any change in $z\sim6$ equivalent width distribution will change 
the $z>7$ values
accordingly.

Following \citet{tre12}, we use the observed $z\sim6$ \wlya\
\citep{sta11} distribution function, $p_6(\wlya)$,  
modeled  as the combination of a Gaussian and a delta function,
\begin{equation}
p_6(W) = \frac{2A}{\sqrt{2\pi} W_c}\exp\left(-\frac{W^2}{2W_c^2}\right) H(W) +
(1-A)\delta(W),
\end{equation} 
where we use the shorthand $W\equiv \wlya$, $W_c=47$~\AA, 
and $A$ is
the fraction  of \lya\ emitters, taken as $A=0.38$ for
$M_\mathrm{uv}<-20.25$ and $A=0.89$ for $M_\mathrm{UV}>-20.25$.  
  $H$
is the Heaviside step function  and $\delta(x)$ is the Delta-function.

\begin{figure}[t]
\epsscale{1.15}
\plotone{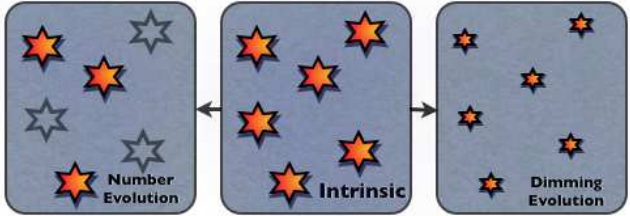}
\caption{
A cartoon representation of our two evolutionary  models$-$ number evolution versus dimming evolution 
used in the Bayesian implementation. In the number evolution scenario only 
some fraction of \lya\ 
emitting galaxies  are visible either due to transparent sight lines or
 other physical 
processes while in the  dimming evolution model, \lya\ photons from all galaxies 
are equally attenuated, say
e.g., due to homogeneous distribution of neutral hydrogen in the
IGM.   In reality,   the observed distribution is likely to be a  combination of both scenarios.
\vspace{0.2cm}
}
\end{figure}

The physics of the evolution of \lya\ emission is likely very
complex, and could involve physical processes associated with the
galaxies that depend on redshift, galaxy mass, inclination angle,
etc., and a highly inhomogeneous (patchy) IGM with rapidly evolving
opacity (see below).     Here, we model the evolution of the probability
distribution $p(W)$ at $z>6$ under two simple empirical cases, as
in Treu et al.\ (2012).  
These two models do not involve any reionization physics but merely 
represent how the observed equivalent width distribution at $z>6$ compares with
 $z\sim6$ distribution.
The first limiting-case is a
\textit{ number  evolution} scenario where only some
fraction, \ep,  of \lya\ galaxies are either completely absorbed or do
not emit \lya\ at all, while  the remaining 1$-$\ep\  are
unattenuated, $p_{ne}(W) = \epsilon_{ne}P_6(W) + (1 -
\epsilon_{ne})\delta(W)$. 
The second limiting-case is  
a uniform \textit{ dimming evolution}
model which could correspond to either evolution in galaxy properties
or a slowly (and homogeneously) evolving IGM neutral fraction (see Fig.~4).
Parametrically, the dimming  model assumes that  \lya\ emission from
all galaxies at $z>6$ is attenuated by the same factor, $\epsilon_{de}$,
such that $p_{de}(W) = p_6(W/\epsilon_{de})/\epsilon_{de}$.  
 Our number evolution and dimming evolution models bracket the range of possible 
 physical effects from reionization and galaxy-evolution physics. For example, 
\citet[][see their Figure 11]{jen14} and \citet{mes14}  show that ÒpatchyÓ reionization  
may  prefer a scenario more similar to our dimming-evolution model.  Regardless, 
the observed equivalent width distribution of \lya\  in galaxies is likely a 
 combination of both scenarios.
These two models are identical in parametrization to the \textit{patchy} and 
\textit{smooth} models of \citet{tre12}. However, 
we avoid using the latter  nomenclature in order
to avoid confusion with the theoretical models that include physics
of reionization. 

While these two models are simplistic,
  they span the full empirical range of evolution in \lya\ emission, which
  allows us to test if the data provide any evidence for the form of the
  empirical evolution.
 Reionization is expected to be a highly patchy process based on theoretical 
 expectations \citep[e.g.,][]{ fur04, ili06, zah07, mes08,  fin09}.
  If  the evolution of 
 the Lyman-alpha emission is a result of such rapid reionization, then 
 we would 
 expect the data to favor the number evolution model.  If on the other hand, the 
 data 
 favors the dimming evolution model, then some other process (or a combination 
 of processes) 
 may be responsible.    
 Therefore, the two  models 
 discussed above  provide a starting point for understanding if there 
 is evidence that
the evolution in the \lya\ emission from galaxies favors ``number'' evolution
versus ``dimming'' evolution.  Furthermore, these models have the
advantage that their effects on the \lya\ distribution function can be
solved analytically, and therefore straightforward to test against
observations.
The models can also be easily adapted as more data, both at $z<7$ 
and at $z>7$, become
available.

%

 \subsubsection{Implementation}

We now  describe how the above  models can be applied to the 
observations.
For an observed  spectrum of a galaxy, the observables are the 
apparent
magnitude $m$, 
the flux density $f_i$ and variance $\sigma_i$
at each pixel $i$ corresponding to each wavelength, $\lambda_i$, 
from the spectroscopic data.
 Following the
methodology in \citet{tre12, tre13},
each wavelength in the
spectrum has some likelihood of containing a \lya\ emission line with
redshift $z_i=\lambda_i /(1216\ \mathrm{\AA})-1$,
 and equivalent
width $W$.  
For an  unresolved line, we model the distribution function of the line flux density as
a Gaussian given by
\begin{equation}
p(f_i,m|W,z_j)=\frac{1}{\sqrt{2\pi}\sigma_i}e ^{-\frac{1}{2}\left (\frac{f_i-W f_m}{\sigma_i} 
\right)^2}
\end{equation}
where
$f_m\equiv f_0\;10^{-0.4m}\; c \lambda_0^{-2}(1+z)^{-1}$,   
$f_0=3.631 \times 10^{-20}$~erg s$^{-1}$ Hz$^{-1}$ cm$^{-2}$.
For a resolved line we replace $Wf_m$ in equation 5 with the line flux distributed as 
a Gaussian with $\sigma_\lambda = 1$~\AA\ (equation 5  in Treu et al. 2012).
%
%
In the region (pixels) where there is no emission line contribution ($W=0$),  
$p(f_i,m|W,z_j)$ is simply  a normal distribution with mean $f_i$ and 
variance
$\sigma_i^2$.   The likelihood of the dataset, $\{f\}$, given a
particular combination of these parameters is then
$p(\{f\},m|\epsilon,z_j)=\int\ \left[\prod_i\ p(f_i,m|\epsilon,z_j)\right]\times 
p(W|\epsilon)\ dW$. 

By Bayes' theorem, the posterior probability on $\epsilon$ is
then simply the product 
 \begin{align}\label{eqn:bayes}
p(\epsilon, z_j|\{f\}, m)=\frac{ p(\{f\},m| \epsilon,z_j)\times  p(\epsilon)\times p(z_j)}{Z},
\end{align}  
where this equation is valid for both $\epsilon=\epsilon_s$ and
$\epsilon_p$.  We adopt a uniform prior $p(\epsilon)$ between 0 and 1
for both cases, where a value of $\epsilon=1$ would imply no evolution.
  The prior on $p(z_j)$ is the photometric redshift 
probability  density 
 for
each galaxy.  Because $p(\{f\}|\epsilon,z_j,m)$ depends on the
S/N ratio of the data, it contains the wavelength-dependent sensitivity 
function. 
The normalization is $Z$=$\prod \left[p(\{f\},m|\epsilon)\right. $$\times$$p(\epsilon)$ 
$\times$ $\left. p(z_j) \right]$.  Intuitively, smaller
values of $Z$ imply that there is less likelihood that the model describes
the data.  The ratio of this factor between two models (number evolution versus 
dimming evolution)  can be used as an 
evidence in favor of one model over the other \citep[e.g.,][]{kas95}.
For the simple models used here,  equation 6 can be solved analytically  \citep{tre12}.

\subsubsection{Results}  
We applied this  Bayesian formalism  to the   spectra of  nine galaxies 
at  $z\sim8$.
In the following we derive the results using these nine spectra 
as well as combining this
sample with that 
 of Treu et al.\ (2013).  
Combining our sample of nine  galaxies with  \citet{tre13}, nearly 
doubles the current 
 spectroscopic sample at $z\sim8$.

\begin{table*}[!t]
\centering
\caption{Results of using different $z\sim6$ equivalent distributions for \ed{our data alone}.}
{\footnotesize 
\begin{tabular}{llcccccc}
  \toprule
Reference &	Distribution type	& Sources used	&  \multicolumn{2}{c}{$\epsilon (W>$25\AA)}  			& \multicolumn{2}{c}{\lya\ fraction ($W>$25\AA)} & Evidence \\
    		&			& 			& 											&								&	relative to $z\sim6$	& & ratio\footnotemark[1]	\\

  {}  	&				& 			& 	$\epsilon_{ne}$	& $\epsilon_{de}$	&	$\rm X_{ne}$	& $\rm X_{de}$			&					  \\
 
   \midrule
St11	& Truncated gaussian plus a 	& $z\sim6$						&  $<$0.56	 & $<$0.74   	& $<$0.56 &$<$0.79						& 2.5 (2.2\footnotemark[2])\\
	& delta function \wlya$>$0\AA	 &  Stark+11, Treu+14			&		&			&    		&							& 	 \\\\

P11	& Same as St11 but with  	& $z\sim6$						&  $<$0.58 & $<$0.75 	&  $<$0.58 &$<$0.80						& 2.3 \\
	& tail extending at 	\wlya$>$100\AA	& Pentericci+11, Treu+12	&  		&		 	& 		&							& 	 \\\\

Sc14	& Log-normal like with 		&  $3<z<6.5$					&  $<$0.59 & $<$0.80 	 &  $<$0.59 &$<$0.86						& 2.4 \\
	& negative tail extending at 	&	as in Schenker et al 2014		&  		&		 	 &		&							& 	 \\
	& \wlya$<$-20\AA		&							&		&		 	 &		&							&	\\\\

T14\footnotemark[3]	& Same as in Sc14 but with	&$-$				& $<$0.59	& $<$0.81		& $<$0.59	&$<$0.86						& 2.3	\\
	& negative tail extending at 	&							&  		&			&    		&							& 	 \\
	& \wlya$<$-50\AA		&							&		&			&		&							&	\\\\

W150\footnotemark[3]	& Same as P11 but with	&$-$				& $<$0.28	& $<$0.18		& $<$0.28	&$<$0.02						& 6.1	\\
	& large number of sources with	&							&  		&			&    		&							& 	 \\
	& \wlya$>$150\AA		&							&		&			&		&							&	\\

\bottomrule
\vspace{-2mm}
\end{tabular}}
\begin{tablenotes}
\item[] These values are for our data only. All limits are 84\% confidence interval.\\
\item[]$\rm ^a$ Bayesian evidence ratio $2\ln (Z_{ne}/Z_{de})$ derived from our dataset alone; see also footnote 9.  \\ 
\item[]$\rm ^b$ The value in parenthesis denotes the Bayesian evidence after combining the posterior for our samples with those of Treu et al. (2013).  \\ 
\item[]$\rm ^c$ This work.  \\ 
  
\end{tablenotes}
\end{table*}

Figure~5  shows the  posterior probability densities derived from nine 
  $z\sim8$ spectra in
our sample, combined with the posteriors taken from
\citet{tre13}.
The  84\% confidence intervals  on  $\epsilon$ are derived by integrating the posterior.
   For our data alone,
we obtain  \ep$<0.56$ and \es$<0.74$  at $z\sim8$ for the number evolution  and 
dimming evolution models 
respectively.   If we combine these results
with those of \citet{tre13},
 we obtain \ep$<0.30$  and \es$<0.25$, 
 for the number  evolution and 
dimming evolution models respectively.

The normalization of the posteriors allow us to derive the Bayesian evidence between
 the two models \citep[e.g.,][]{jef61, kas95}\footnote{
 Kass \& Raftery (1995) define  the  significance  scale in favor of one model over the other 
using  the Bayes factors $Z_1$ and $Z_2$ as  $S=2 \ln(Z_1/Z_2)$, with S=$0-2$ (not worth
more than a bare mention), S=$2-6$ (positive), 
S=$6-10$ (strong), S$>10$ (very strong).}.
 For our data alone, we find a  tentative ``positive"  Bayesian evidence favoring 
 the number evolution model, with $2\ln (Z_{ne}/Z_{de}) = 2.5$.  This evidence drops slightly 
but remains ÒpositiveÓ towards the number evolution model with  $2\ln(Z_{ne}/Z_{de}) = 2.2$ when we 
combine our data with that from Treu et al. (2013).  Therefore the evidence is minimally 
significant that the evolution in the \wlya\ distribution at $z\sim 8$ favors the number evolution model.

\begin{figure}[t]
\epsscale{1.2}
\plotone{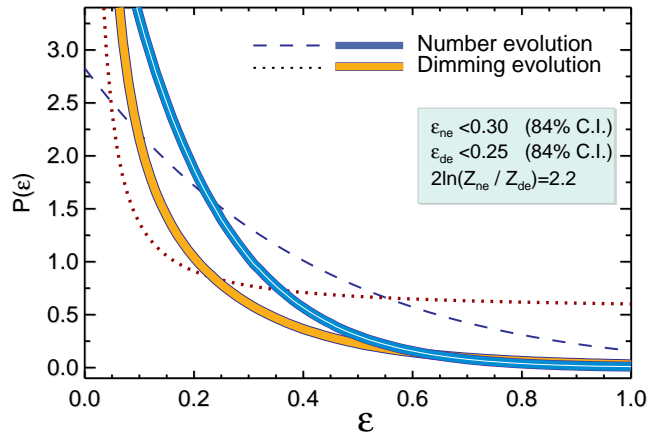}
\caption{
Posterior probability density  of $\epsilon$ where
$\epsilon$  is the proportionality factor  between \lya\ fraction 
$X_{z\sim8}$  and $X_{z\sim6}$.
Blue and orange  lines show the probability using the combined sample (This work + Treu 
et al 2013),
for  number evolution and  dimming evolution  models, respectively.
Also labeled are the  84\% value for  both models.
Dashed and dotted lines represent probability density using our dataset alone. 
The data provide ``positive'' evidence (2$\,$ln$(Z_p/Z_s) =2.2$) in favor of the  number evolution model
over the  dimming evolution model using the significance scale of \citet{kas95}.
\vspace{0.2cm}
}
\end{figure}

Qualitatively, the reason the Bayesian evidence favors number evolution  of the 
\wlya\ distribution is that even the relatively small sample sizes are becoming 
large enough to discriminate between these simple evolutionary models.  
For example, under the dimming evolution model, the  \wlya\ distribution shifts such that 
there are relatively many  objects with low \wlya\ and  fewer objects 
with high \lya, while the total number of objects with \lya\ emission remains unchanged
(assuming one can detect the lower levels of \lya\ emission).   
 If the true evolution follows the dimming evolution model, then we would expect  more 
 \lya\ detections at low \wlya, which is not observed and is therefore less favored by 
 the Bayesian analysis.     In contrast, in the number evolution model,   some fraction of \lya\ 
 sources are blocked, keeping the relative distribution of \lya\ unchanged, and 
 is favored by the current data.     
 Clearly larger sample  will be needed to 
 confirm these results and/or  increase the evidence against the dimming evolution model.

The posterior distribution of $\epsilon$ is broader for our sample compared to that of 
\citet{tre13},
 because we detect  \lya\ in one object \citep{fin13}
 and 
have three other marginal detections ($\simeq 2-3\sigma$). 
In addition, our MOSFIRE Y-band
observations  did not fully cover $p(z)$ of all nine galaxies.
Thus, there is a finite probability that the \lya\ line
could lie outside the  Y-band, which again  broadens the $\epsilon$ distribution.

\section{Discussion}

Based on the  $\epsilon$ constraints above, and 
fitting functions used for $z\sim6$ equivalent width
distributions, the \lya\ fraction in the number evolution  model is
$X_{z\sim8}=\epsilon_p X_{z\sim6}$, while  in the dimming evolution case
$X_{z\sim8}=$ erfc$(W/\sqrt{2}\epsilon_s W_c)$$/$ erfc$(W/\sqrt{2} W_c)
 X_{z\sim6}$ where \textit{erfc} is the complimentary error function.

For our dataset alone, the \lya\ fraction $X_{\rm z\sim8}<0.56X_{\rm z\sim6}$ and 
$X_{\rm z\sim8}<0.79X_{\rm z\sim6}$ (all 84\% confidence limits) for number evolution and dimming evolution models
respectively.
For the combined data (with Treu et al 2013), 
$X_{\rm z\sim8}<0.30X_{\rm z=6}$ and 
$X_{\rm z\sim8}<0.05X_{\rm z=6}$ for number evolution  and dimming evolution models
respectively.
In Figure~3 we show  the  constraints from the number evolution model only (as
this is the conservative limit)  as filled blue circles.  For the
brighter  sample, $X_{\rm z\sim8}<0.06$ and $X_{\rm z\sim8}<0.01$ while
for the fainter sample $X_{\rm z\sim8}<0.16$ and $X_{\rm z\sim8}<0.03$
for number evolution and  dimming evolution  models respectively.  The implication is that
at $z>7$, the fraction of \lya\ emitters is reduced by a factor   of
$>3$ (84\% confidence interval) compared to the fraction at $z\sim6$.
At the 95\% confidence interval, the \lya\ decline at $z>7$ is $>2$, implying a strong 
evolution even at this more conservative limit.

\subsection{Effect of Using Different $z\sim6$ Distributions}

Our results show that there is  strong evidence for evolution in the 
Lyman-$\alpha$ equivalent width distribution.  The nature of this 
evolution however depends  on the assumed   $z=6$ equivalent width
distribution \citep{sta11}.
Here we test other possible $z=6$ Lyman-$\alpha$ 
equivalent width distributions to see how this choice affects our conclusions.

To test this effect, \citet{tre12} explored $z\sim6$ 
equivalent width distribution with a tail extending towards larger equivalent width objects,
similar to \citet{pen11},  for fainter sample. 
They find that this equivalent width distribution
does not alter their conclusions.
We performed a similar test on our data using $z\sim6$ equivalent width distribution with a
uniform  tail extending towards higher$-$\wlya\ (150\AA) objects.
Using this equivalent width distribution, the  Bayesian probabilities change only slightly with 
\ep\ changing from \ep$<0.56$ to \ep$<0.58$ and \es\ increases from 
\es$<0.74$ to \es$<0.75$ for our data.

 \begin{figure}[h!]
\epsscale{1.19}
\plotone{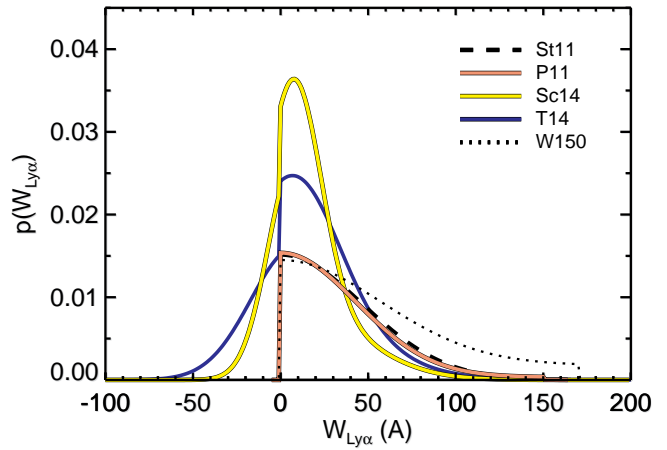}
\caption{
\ed{
Different $z\sim6$ \lya\  equivalent width distributions used to test their effect on our results. Legends
indicate different distributions described in Table 1.
The dotted line (W150) is an arbitrary distribution chosen to demonstrate how different the $z\sim6$ equivalent 
distribution needs to be in order to significantly change the \lya\  fraction evolution from the current 
value. This distribution yields much lower \lya\ fraction   with
$X_{\rm z\sim8}<0.28X_{\rm z=6}$ for the number evolution model.}
\vspace{0.2cm}
}
\end{figure}

In addition, we explored two more  $z\sim6$ equivalent width distributions: 1) similar
to the log-normal distribution used in \citet{sch14} with additional tail extending towards
negative equivalent widths (\wlya$< -20$\AA) and 2) an extreme distribution with negative
tail extending to   \wlya$< -50$\AA. We show these distributions in Figure~6 and 
 tabulate the results of these tests  in Table 1, using 
our data alone as the results for these distributions are not available from \citet{tre13}.
Thus, it can be seen that our results do not change significantly if the input $z\sim6$ distribution
includes a significant tail extending to very negative equivalent widths.
\ed{
However, if  the $z\sim6$ \lya\ distribution is significantly different, for example the one similar to the
dotted line (W150) shown in Figure~6, then the derived constraints would be significantly different. 
The W150  
 distribution which  contains large number of galaxies with high   \wlya$> 150$\AA\ 
yields $X_{\rm z\sim8}<0.28X_{\rm z=6}$ 
for the number evolution model, much lower compared to the other input equivalent width distributions.
Thus, it is important to have a much larger spectroscopic sample of $z\gtrsim 6$ galaxies in order to 
construct a robust equivalent width distribution at this redshift.
}

\subsection{Physical Interpretation}
There are a few possible reasons that can explain the observed  sharp decline
in the \wlya.  It may be the samples are biased:  we may be
preferentially  targeting only bright continuum galaxies with lower
$W_{\mathrm{Ly}\alpha}$. However, this is unlikely because Figure~1
shows that the galaxies with Lyman-$\alpha$ span a wide range of
$M_\mathrm{UV}$ and that UV continuum limits of $z\gtrsim7$ surveys 
 are nearly same as that
of  $z<7$ surveys.    It may also be that the $z>7$ samples include a
larger number of lower-$z$ contaminants.  We think this is unlikely
because it does not explain the steep decline in the \wlya\  values (top panel in Figure~1)
 in spectroscopically detected galaxies at $z>6.5$,

 It may be  evolution in the intrinsic galaxy properties that is
responsible.  
Previous work has shown that galaxies at
$z>6$  have lower dust extinction \citep[e.g.,][]{fin12}, smaller
sizes  and lower stellar mass, making it easier 
 for Lyman-$\alpha$ photons to escape.   \citet{rob10} 
 also find evolution in the UV continuum properties of galaxies, and 
 discuss how this may contribute to the decline in the \wlya\ in 
 spectroscopic samples.  Galaxies at $z > 7$ may also have higher 
 gas accretion rates relative to their SFRs than lower redshift galaxies 
 \citep[e.g.,][]{pap11, fin12}, which could 
 suppress the escape of Lyman-$\alpha$ photons depending on the 
 gas dynamics (infall velocity versus outflow rates) or if the covering 
 fraction of the infalling gas is large.  
  On the other hand, recent work 
\citep[e.g.,][]{iwa09} implies that the Lyman-continuum photon 
  escape fraction should be \textit{higher} at $z>3$ than at lower redshift to account for
reionization,  but see also \citet{bou11} where they find a low continuum escape
fraction in Lyman-break selected galaxies at $z=3.3$.

While some evolution of the physical properties of $z >7$ galaxies occurs,
  it seems galaxy evolution alone is unable to account for the decline in the 
  Lyman-$\alpha$ distribution.   Rather,  the most plausible explanation is that 
  both a changing neutral hydrogen fraction in the IGM and evolution from 
  galaxy properties contribute to the decline in the \wlya\ distribution.
%
%
Based on our
results,  the fraction of galaxies with strong \lya\ equivalent widths
has dropped significantly from $z\sim6.5$ to $z>7$.
If this  suppression in \lya\ emission is solely due to the IGM evolution, it
can be directly interpreted as an
increase in the optical depth, $\Delta \tau_{\mathrm{Ly}\alpha}$, from
$z=6.5$ to $z=7.5$, 
where $\langle \exp(-\Delta \tau_{\mathrm{Ly}\alpha}) \rangle$ =
$\epsilon$ (for both the number and dimming evolutionary models).  
Based on our results, in a simplistic case where  reionization is
uniform, the conservative limit  on the optical depth of $\Delta
\tau_{Ly\alpha}>1.2$ (84\% confidence interval) at $z\sim7.5$.  This is similar to
the rapid evolution in the opacity of the \citet{gun65}
trough, where the extrapolated relation from \citet{fan06}
predicts $\Delta \tau=(1+z_{7.5})/(1+z_{6})^{4.3}=2.3$.

The observed rapid decline in the \lya\ fraction 
is consistent with recent theoretical predictions. However, 
these studies suggest that at $z\sim7$ 
we need only $\sim 10-20\%$ neutral hydrogen fraction  to explain the observed
\lya\ fraction decline if we account for 
 cosmic variance, evolving escape fraction of ionizing photons and 
increasing incidence of optically thick absorption systems 
\citep{bol13, tay13, dij14}.
\citet{mes14} argue based on their model that, at 68\% confidence, the decline in the Lyman-$\alpha$ fraction at $z\sim 7$ from lower redshift can not be greater than a factor of two unless galaxy evolution processes also contribute to the decline in Lyman-$\alpha$ photons.  However, at 95\% confidence the observed decline in Lyman-$\alpha$ may stem solely from the evolution of the IGM in their model.
At $z>7$, however, it is possible that there is in fact a more rapid increase
in the neutral hydrogen fraction which may explain the steep
observed decline of \lya\ equivalent widths of
individual galaxies  (see \S 4.3 below). 
To investigate and minimize these observational uncertainties, we therefore need much 
larger spectroscopic
samples of $z>7$ galaxies.

In addition to the declining \lya\ fraction, our simplistic models show
first tentative evidence towards  number  evolution scenario at $z>7$,
extending recent results seen at $z\lesssim7$ \citep{pen14}.
 Therefore, the data support the idea that some process
 related to decreasing high \wlya\ galaxies  is dominant.  This is consistent with
 reionization where regions are opaque, likely due to neutral hydrogen, blocking a
  fraction of sightlines while leaving others unaffected.   This makes
  the prediction that if we survey enough area, we should find objects
  with high \wlya, but they should be rare.  Indeed, there is a 
  recent report of a 
  weak detection ($\sim4\sigma$) of a galaxy at $z = 7.6$ with
  \wlya $=160$\AA\ \citep{sch14}, which if confirmed could further support
  our interpretation. \\

\subsection{Implications for Reionization}

Several theoretical studies using semi-analytical and numerical simulations have developed 
models of IGM evolution and its effect on the observations 
\citep[e.g.,][]{mir00, cia06, gne06, fur06, mcq07, mes08, cho09, cro11, dij11, alv12, jen13}.
In this section, to estimate the neutral hydrogen fraction \fhi\ at $z\sim8$,   we use two 
different  models which predict the  probability of  \lya\
equivalent widths given certain neutral hydrogen fraction in the IGM combined with line-of-sight
\lya\ absorbers \citep{bol13} and models that include  evolving escape fraction 
of ionizing photons \citep{dij14}.

Figure~7 shows the cumulative probability distribution of \lya\ equivalent widths 
comparing our
results  with the theoretical predictions from \citet{bol13} and \citet{dij14}
for the fainter sample.
Our results are shown only for the number evolution model.
The Cyan-filled region shows model predictions at $z\sim7$ from \citet{bol13} for
a range of \lya\ velocity offsets from 200 to 600 km s$^{-1}$,  photo-ionization rate
log$(\Gamma_{\rm HI}/S^{-1})=-14$, and volume average neutral hydrogen fraction of 
\fhi$\sim 0.1$.
The yellow-filled region shows the model prediction at $z\sim8$ \citep{dij14} for 
\fhi$\sim0.3$ and 
escape fraction of ionizing photons
$<f_{esc}>=0.04 [(1+z)/5]^4$.
Compared with these model predictions, our current \lya\ emitter fraction  are lower by a 
factor of $\sim2$.
Thus, we conclude that the neutral hydrogen fraction at $z\sim8$ is \fhi$\gtrsim0.3$
considering the evolution of neutral hydrogen fraction as well as evolving galaxy
properties such as winds, ionizing escape fraction, etc.
If the decline is solely due to the reionization, the amount of neutral hydrogen 
fraction at $z\sim8$ will be much higher because the model predictions \citep{bol13, dij14} 
in Figure~7 already include some galaxy evolution.
This is consistent with inferences of the neutral hydrogen fraction based on the  
evolution of the UV and \lya\ luminosity functions \citep{rob13, kon14}, and thus,
the reionization of the universe is likely in progress  at $z\sim8$.

\begin{figure}[t]
\epsscale{1.17}
\plotone{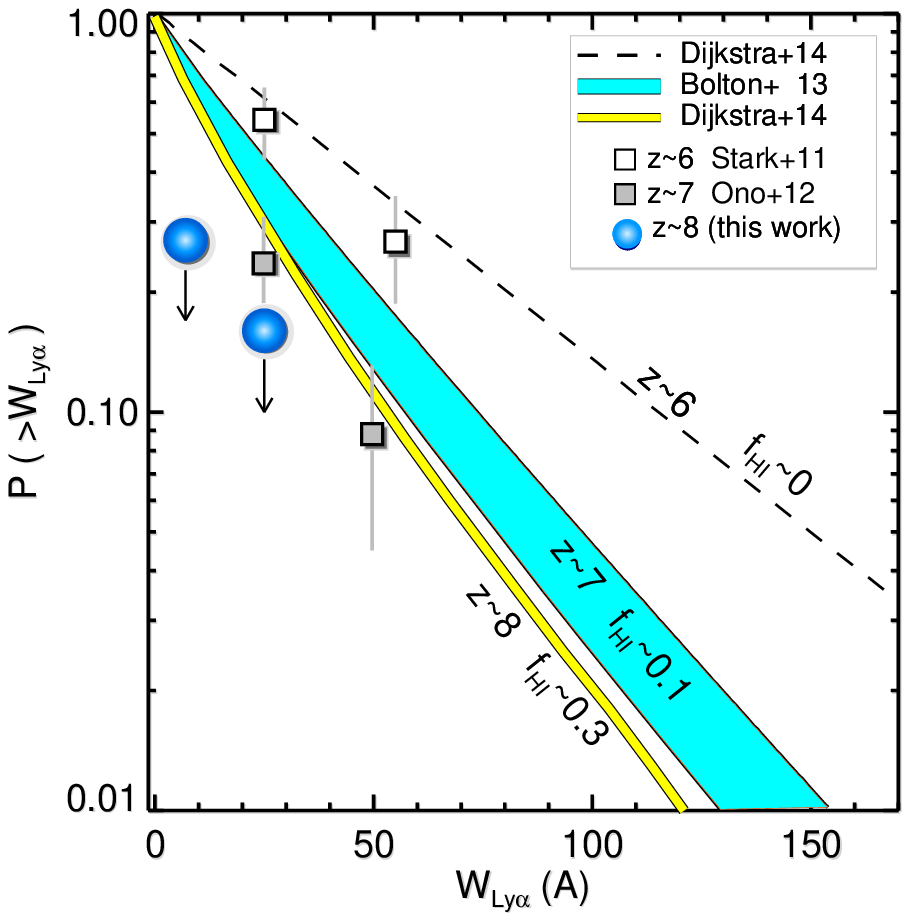}
\caption{
Cumulative \lya\ probability distribution  for the faint sample (\muv$>-20.25$ mag) and comparison of model predictions 
\citep{bol13, dij14}
 to estimate the
 neutral hydrogen fraction in the IGM at $z\sim8$. 
Dashed line shows the reference $z\sim6$ \lya\ EW distribution  
used in the models, equivalent to a neutral 
hydrogen fraction \fhi$\sim0$ while open squares represent $z\sim6$ observations from \citep{sta11},
used in our Bayesian models.
Cyan-filled regions shows model predictions at $z\sim7$
for a volume averaged \fhi$\sim0.1$
(Bolton et al 2014). Filled Gray squares are the observations from \citep[][and references 
there in]{ono12}.
The yellow-filled region  shows the model  prediction at $z\sim8$ \citep{dij14} 
for \fhi$\sim0.3$.
Our  results, shown in filled blue circles (shown only for the  number evolution scenario)
 are significantly lower compared with the model 
predictions. This implies that  the volume-averaged neutral 
hydrogen fraction at $z\sim8$ is at least  \fhi$\gtrsim 0.3$.
\vspace{0.0cm}
}
\end{figure}

\section{Summary}

We investigated the evolution of the \lya\ fraction from $z\sim6$ to $z\sim8$ using 
extremely
deep spectroscopic observations of nine galaxies, obtained using the MOSFIRE Y-band that 
covers
the redshift range $7<z<8.2$.
We explored two different methods to study the \lya\ fraction : a direct method with
extensive completeness simulations to account for the incompleteness and a
Bayesian inference method using two simplistic models$-$ number evolution  versus dimming 
evolution.

The Bayesian method yields much stronger constraints than the direct method 
due to its 'relative' inference$-$ the \lya\ fraction at $z\sim8$ is relative to the $z\sim6$ 
values
and any change in $z\sim6$ values will change the derived $z\sim8$ values accordingly.
Combining our data with that of \citet{tre13}, we found that the \lya\ fraction at $z\sim8$
has dropped significantly, by a factor of $>$3 (84\% confidence), 
 compared with $z\sim6$ values. 
However, it may be that the other factors such as (rapid) evolution in galaxy properties, or 
field-to-field variations also affect the \lya\ emission distribution.

Our results show a  tentative ``positive" evidence towards the number evolution model with Bayesian evidence 
ratio of 
2$\,$ln$(Z_{ne}/Z_{de}) =2.2$ extending earlier $z\sim7$ results to higher redshift,  $z>7$.
Furthermore, comparing our results with theoretical predictions, we find that the neutral
hydrogen fraction \fhi\ at $z\sim8$ is $\gtrsim0.3$.
To corroborate these results further, and understand how the Lyman-$\alpha$ width 
distribution function evolves from $z=6$ to $z>7$,  we need 
 larger samples of galaxies (particularly with high \wlya).
 Only with that
knowledge can we constrain the nature of reionization.

\acknowledgements
We thank Tommaso Treu and Kasper Schmidt for useful conversations and comments
on the manuscript. We also thank James Rhoads and Sangeeta Malhotra for helpful 
comments and suggestions.
The authors also thank the referee for a careful reading,  and for comments and 
suggestions that improved the quality and clarity of this work.
The authors wish to recognize and acknowledge the very significant cultural role and 
reverence that the summit 
of Mauna Kea has always had within the indigenous Hawaiian community. We are most 
fortunate to have the 
opportunity to conduct observations from this mountain.
We also thank  the CANDELS team  for providing a valuable dataset.
This work is supported by HST program GO-12060,
provided by NASA through a grant from the STScI, which is operated by the AURA.

{}

\end{document}